\documentclass[prd,preprint,superscriptaddress]{revtex4}
\usepackage{revsymb}
\usepackage{amsmath}
\newcommand{\be}{\begin{equation}}
\newcommand{\ee}{\end{equation}}
\newcommand{\ben}{\begin{eqnarray}}
\newcommand{\een}{\end{eqnarray}}

\newcommand{\vphi}{\varphi}
\begin{document}
\title{Exact solutions for the interacting tachyonic--dark matter system}
\author{Ram\'{o}n Herrera\footnote{E-mail address: ramon.herrera.a@mail.ucv.cl}}
\affiliation{Instituto de F\'{\i}sica, Pontificia Universidad
Cat\'{o}lica de Valpara\'{\i}so, Avenida Brasil 2950, Casilla
4059, Valpara\'{\i}so,Chile}
\author{Diego Pav\'{o}n\footnote{E-mail address: diego.pavon@uab.es}}
\affiliation{Departamento de F\'{\i}sica, Facultad de Ciencias,
Universidad Aut\'{o}noma de Barcelona, 08193 Bellaterra
(Barcelona), Spain}
\author{Winfried Zimdahl\footnote{E-mail address: zimdahl@thp.Uni-Koeln.DE}}
\affiliation{Institut f\"ur Theoretische Physik, Universit\"{a}t
zu K\"{o}ln, 50937 K\"{o}ln, Germany}
\begin{abstract}
We find exact solutions leading to power law accelerated expansion
for a homogeneous, isotropic and spatially flat
universe, dominated by an interacting mixture of cold dark matter
and a tachyonic field such that the ratio of the energy
densities of both components at late times is constant 
and no coincidence problem arises.
\end{abstract}
\maketitle
\section{Introduction}
Recently the tachyon field introduced by Sen in a series of papers
\cite{sen}, has attracted some attention in cosmology
\cite{Gibbons,Pad,Fair,Fein,Chaud,Frolov,Kof,Wasserm, Gorini,Abramo}.  As
shown by Bagla {\it et al.} \cite{bagla} the Lagrangian of the tachyon
field $L = - V(\vphi) \sqrt{1 - \partial^{a} \vphi \, \partial_{a} \vphi}$
arises as a straightforward generalization of the Lagrangian of a
relativistic particle, $L = -m \sqrt{1 -\dot{q}^2}$, much in the
same way as the Lagrangian of the scalar field generalizes the
Lagrangian of a non--relativistic particle. Likewise, its stress
energy tensor has the structure of a perfect
fluid and it can be seen as the sum of dust matter and a cosmological
constant whereby it may play the role of both dark matter and
dark energy \cite{Chaud,bagla}. This latter feature together with the
fact that the pressure associated to the tachyon field is negative
-a key ingredient in Einstein's relativity to produce 
accelerated expansion-  may
explain the interest in using it to account for the
present state of the Universe \cite{expansion}.
Further, it has been argued that tachyonic fields may describe a
larger variety of interesting physical situations than quintessence fields
\cite{Gorini}. Besides,  when the tachyon field potential reduces
to a constant, its equation of state coincides with that of the
Chaplygin gas -see, e.g., \cite{chaplygin}.

To study the cosmological dynamics in the presence of tachyonic
matter, several authors have resorted to the potential $V (\vphi)
\propto \vphi^{-2}$  (where $\vphi$ is the tachyon field), since
it leads to a power--law solution for the scale factor of the
Robertson--Walker metric.  (Bear in mind that in the case of a
minimally coupled scalar field a power law behavior requires an
exponential potential.) Here we are interested in situations where
the cosmic medium may at present be regarded as a mixture of
two components, namely a tachyon field (acting as dark energy)
and pressureless dust (i.e., cold dark matter) such that the ratio
between their energy densities is a constant, thus indicating a
solution of the coincidence problem which afflicts many
approaches to late acceleration. (Obviously, this makes sense
only, if the tachyon field is not yet near its long--time limit in
which its equation of state approaches that for dust as well).
This is accomplished by assuming an interaction between both
components so that they do not conserve separately as the Universe
expands. This approach parallels somewhat a previous study in
which a generic quintessence scalar field interacted with cold
dark matter in such a way that the resulting dynamics was
compatible with late acceleration and solved the coincidence
problem  \cite{plb}.

We find exact solutions to the system of dark matter and tachyon
field equations for different choices of $V(\vphi)$ such that the
ratio between the energy densities of both components is kept
constant  at late times  and the scale factor of the 
Robertson--Walker metric obeys a power law.

The next section presents the relevant field equations of the
tachyonic field. In section III the interaction between the latter
and cold dark matter is considered and some solutions are found,
first when $\dot{\vphi}^{2}$ is held constant and then when this
constraint is relaxed. 
Finally, section IV summarizes our findings.

\section{Basic tachyon field equations}
We begin by succinctly recalling the basic equations of the tachyon field
to be used in the next section where the interaction with dark matter
is incorporated.

The stress--energy tensor of the tachyon field
\\
\begin{equation}
T^{(\vphi)}_{ab} = \frac{V\left(\vphi\right)}
{\sqrt{1+{\vphi}^{,c} {\vphi}_{,c}}}
\left[-g_{ab}\left(1 +{\vphi}^{,c} {\vphi}_{,c} \right) +
\vphi_{,a}\vphi_{,b}\right] \ ,
\label{stress1}
\end{equation}
\\
admits to be cast into the form of a perfect fluid
\\
\begin{equation}
T^{\vphi}_{ab} = \rho_{\vphi} u_au_b + p_{\vphi} \left(g_{ab} +
u_au_b\right),
\label{stress2}
\end{equation}
\\
where the energy density and pressure are given by
\\
\begin{equation}
\rho_{\vphi}  = \frac{V(\vphi)}{\sqrt{1-\dot{\vphi}^2}}
\qquad \mbox{and} \qquad
p_{\vphi}  = - V(\phi){\sqrt{1-\dot{\vphi}^2}},
\
\label{rhop}
\end{equation}
\\
respectively, with
\\
\begin{equation}
\dot{\vphi } \equiv  \vphi_{,a}u^{a}
= \sqrt{-g ^{ab}\vphi_{,a}\vphi_{,b}} \quad \mbox{and}
\quad u_{a} = - \frac{\vphi_{,a}} {\dot{\vphi}}\ ,
\quad \mbox{with} \quad
u ^{a}u _{a} = -1.
\label{respectively}
\end{equation}

Furthermore if the tachyon field interacts only gravitationally,
its evolution equation reads
\\
\begin{equation}
\frac{\ddot{\vphi}}{1-\dot{\vphi}^2} + 3H \dot{\vphi}
+ \frac{V'}{V}= 0 \, ,
\label{consv2}
\end{equation}
\\
where $H \equiv \dot{a}/a$ is the Hubble factor and $a$ the scale factor
of the FLRW metric; the prime indicates derivation respect to $\vphi$.
The latter expression can be recast as
\\
\begin{equation}
\dot{\rho}_{\vphi}= - 3 H\dot{\vphi}^2\rho_{\vphi}\, ,
\label{consv1}
\end{equation}
\\
this implies that for any $\dot{\vphi}^2 < 1$ the energy
density of tachyon field decays at a lower rate than that for
dust. It approaches the behavior of dust for $\dot{\vphi}^{2}
\rightarrow 1$.  In this limit the tachyon may be considered a
pressureless dark matter component.

As mentioned above, we shall assume that in addition to the tachyon
field (with $\dot{\vphi}^2 < 1$) a cold dark matter fluid, of energy
density $\rho_{m}$, enters the cosmic medium. Thus, the Friedmann
equation for this two component system in a spatially flat FLRW
universe can be written as
\\
\begin{equation}
H^2 = \frac{8\pi G}{3}\left[\frac{V\left(\vphi\right)}
{\sqrt{1-\dot{\vphi}^2}} + \rho_{m}\right]  \ .
\label{friedmann}
\end{equation}

\section{The $\vphi$CDM interacting model}
Henceforward we shall assume that both components -the tachyon
field and the cold dark matter- do not conserve  separately
but that they interact through a term $Q$ 
(to be specified later) according to
\\
\begin{equation}
\dot{\rho}_m + 3H \rho_m = Q \ ,
\label{interact1}
\\
\end{equation}
\begin{equation}
\dot{\rho}_{\vphi} + 3H \dot{\vphi}^2 \rho_{\vphi} = -Q \ .
\label{interact2}
\end{equation}
\\
The interaction term $Q$ is to be determined by the condition that the
ratio between the energy densities $r \equiv \rho_{m}/\rho_{\vphi}$
remains constant at late times. One easily realizes that a 
suitable interaction between both components, 
\\
\begin{equation}
Q = 3H \frac{r}{(r+1)^{2}}(1-\dot{\vphi}^{2}) \rho \ ,
\label{Q1}
\end{equation}
\\
where $\rho = \rho_{m} + \rho_{\vphi}$ is the total energy density of the
cosmic substratum,  leads to the desired result. 
Since $\dot{\vphi}^{2} < 1$ we have $Q > 0$. Therefore, for a
stationary ratio $r$ to exist, a transfer of energy from the
tachyon field to the matter component is required.  A stability
analysis of the stationary solution parallel to that in \cite{plb}
reveals that when $Q /3H \propto \rho$ in the vicinity of the
stationary solution, then $r$ is stable for any $r <1$ (and
$\dot{\vphi}^2 < 1$). In particular, the stability is compatible
with accelerated expansion (see below). In the presence of the
above interaction the evolution equation for $\vphi$ (Eq.
(\ref{consv2})) generalizes to
\\
\begin{equation}
\frac{\ddot{\vphi}}{1-\dot{\vphi}^2} + 3H \dot{\vphi}
+ 3H \frac{1-\dot{\vphi}^2}{\dot{\vphi}}\frac{r}{r+1}
+ \frac{V'}{V}= 0 \ .
\label{ddvphi1}
\end{equation}

\subsection{$\dot{\vphi}^2 = {\rm const}$}
When $\dot{\vphi}^2 = {\rm const}$, the solution
\\
\begin{equation}
\rho_m \ , \rho_{\vphi}\ , V \propto a^{-3\frac{r +
\dot{\vphi}^2}{r+1}} \, ,
\label{rhoV}
\end{equation}
\\
readily follows. It corresponds to a power law expansion
\\
\be
a(t) \propto t^{n} \, , \qquad {\rm with}  \qquad
n = \frac{2}{3}\frac{r+1}{r +\dot{\vphi}^2} = {\rm constant} \ .
\ee
\\
Thus, the temporal evolution is given by
$\rho_{m}, \rho_{\vphi} \propto t^{-2}$.
Thereby, one has accelerated expansion for
\\
\begin{equation}
n > 1 \quad\Longleftrightarrow\quad \dot{\vphi}^2 < \frac{2 - r}{3}\ ,
\label{22}
\end{equation}
\\
i.e., $\dot{\vphi}^2$ has to be sufficiently small.
For $\rho_{m} = r = 0$ one recovers the result of 
the single--component case
\cite{Gibbons}.

\subsection{$\dot{\vphi}^2 \neq {\rm const}$ }
The solution just found is a very particular one as, generally
speaking, one should expect that $\ddot{\vphi}$ does not vanish.
Therefore, we next focus on finding analytical solutions such that
$\ddot{\vphi} \neq 0$ but, as before, keeping the ratio $r$
between the energy densities constant.

It seems unclear if this can be accomplished at all by 
a perfect (i.e., non--dissipative) fluid model of the dark matter. 
But as we shall see, this can be achieved
by assuming that the dark matter component is dissipative, i.e.,
endowed with a dissipative pressure $\pi_{m}$ whose origin may lie
either in the interactions between the particles that make up the
dark matter fluid or in their decay and/or mutual annihilation
-for a recent short review on models of self--interacting CDM see
\cite{ostriker}. As a consequence, the energy balance equation for
the matter component now reads
\\
\begin{equation}
\dot{\rho}_m + 3H (\,\rho_m \,+\,\pi_{m}\,)= Q \ ,
\label{consv3}
\end{equation}
where the strength of the interaction $Q$ will differ from that in
Eq.(\ref{Q1}). The presence of $\pi_{m}$ (which ought to be
negative for expanding fluids as required by the second law of
thermodynamics -see, e.g., \cite{landau-lifshitz}) in the dark
matter fluid is only natural since (barring superfluids) all
matter fluids found in Nature are dissipative -see e.g.
\cite{batchelor}. Further, this quantity is crucial to solve the
coincidence problem of late acceleration in models where the dark
matter and dark energy conserve separately \cite{key}. Here we
shall use the imperfect fluid degree of freedom to obtain a
power-law solution for $\dot \vphi\neq$ constant under the
condition $r =$ const.

The energy balance equation for the tachyon field can be written as
\\
\begin{equation}
\dot{\rho}_{\vphi} + 3H \dot{\vphi}^2 \rho_{\vphi} = - Q \ ,
\label{consv4}
\end{equation}
\\
or equivalently,
\\
\begin{equation}
\frac{\ddot{\vphi}}{1-\dot{\vphi}^2} + 3H \dot{\vphi} +
\frac{V'(\vphi)}{V(\vphi)}= -
\frac{\sqrt{1-\dot{\vphi}^2}}{V(\vphi) \dot{\vphi}} \, Q \ .
\label{consv5}
\end{equation}

Again, we have left the interaction term unspecified. To determine
it we impose (as we did above) that the ratio of both energy densities
remains constant (i.e., we demand that the coincidence problem
should be solved).  Thus, in addition to equations (\ref{consv3})
and (\ref{consv4}) we require that
$\dot{r} =(\rho_{m}/\rho_{\vphi})^{\displaystyle \cdot}$ = 0. 
As a consequence,
the interaction term is now given by
\\
\begin{equation}
Q = 3H \frac{r}{\left(r + 1\right)^2} \left(\frac{\pi_m}{\rho_m} -
\frac{p_\vphi}{\rho_\vphi}\right) \rho \ . \label{}
\end{equation}
This corresponds to
\\
\begin{equation}
\frac{\dot{\rho}_m }{\rho_m} = \frac{\dot{\rho}_{\vphi}
}{\rho_{\vphi}} = -3H \left[1 + \frac{p_\vphi +
\pi_m}{\rho}\right] = - 3H \left[1 - \frac{1 - \dot\vphi ^2}{1 +
r} + \frac{\pi_m}{\rho}\right]\ . \label{corresponds}
\end{equation}
\\
For the expression in the brackets on the right hand side to be a constant when 
$\dot \vphi ^2$ is admitted to vary, the last term must cancel 
the $\dot \vphi ^2$ term.
This suggests an ansatz
\\
\begin{equation}
\pi = - b^2 \rho \ , \label{}
\end{equation}
\\
where
\\
\begin{equation}
b^2 = b_0 ^2 + \frac{\dot \vphi ^2}{r+ 1} \ ,\label{}
\end{equation}
with $b_0^2 =$ const, which implies
\\
\begin{equation}
\frac{\dot{\rho}_m }{\rho_m} = \frac{\dot{\rho}_{\vphi}
}{\rho_{\vphi}} = -3H \left[\frac{r}{r + 1} - b_0^2\right] \ .
\label{}
\end{equation}
\\
Physically, this means that the ratio of the total pressure, which
is $p_\vphi + \pi_m$, to the total energy density $\rho$ is required not
to depend on $\dot\vphi^{2}$. In other words, $\pi_m$ has to be
chosen such that the total equation of state of the cosmic medium
is constant, while at the same time $\dot \vphi^{2}$ is not. Thus,
the dependences of $\rho_m$ and $\rho_{\vphi}$ on the scale 
factor are found to be:
\\
\begin{equation}
\rho_m\propto\, a^{-\nu
},\,\,\,\rho_{\vphi}\propto\,a^{-\nu},\,\,\,\,\,\,
\nu=3\left[\frac{r}{r+1}-b^2_o \right]\ .
\label{dependence}
\end{equation}
\\
Under this condition the interaction term is now given by
\\
\begin{equation}
Q = 3H \frac{r}{\left(r + 1\right)^2} \left[1
 - \frac{1 + r}{r}\left(b_0^2 + \dot \vphi^2\right)\right] \rho
 \ .
\label{interaction}
\end{equation}

By virtue of the relationship $\rho\propto\,a^{-\nu}$ the
Friedmann equation (\ref{friedmann}) leads to $a(t)
\propto\,t^{2/\nu}$. It readily follows that
$\rho\propto\rho_m\propto\rho_{\vphi}\,\propto\,t^{-2}$ for the
energy densities. Again, for a power--law solution to exist, a
transfer of energy from the tachyon field to the matter component
is required (i.e., one must have $Q > 0 $), as well as $\nu <2$
(tantamount to $r/(r+1)<(2/3)+b^2_o$), to have accelerated
expansion.

We conclude that by a suitable choice of the imperfect fluid
degree of freedom $\pi_m$ it is indeed possible to obtain a power
law solution with $r =$ constant and $\dot\vphi \neq$ const. The
magnitude of this pressure is  largely dictated by the dynamics of the
tachyon field. Since from the outset $\pi_m$ was introduced to
account for interactions within the matter component, this may
seem surprising at a first glance. The point here is that the
requirement $r =$ constant establishes a strong coupling of the
dynamics of the components which are no longer independent of each
other. Only if $\pi_m$ is such that $\pi_m /\rho$ follows the (not
yet known) dynamics of $\dot \vphi ^2$, the described power-law
solution is possible. As we shall see, such a configuration may
serve as a toy model which admits exact solutions for the
cosmological dynamics.

A stability analysis of the stationary solution may be performed
as in \cite{plb} where the relationship $Q = 3H c^{2} \rho$,
with $c^{2}$ a constant, was hypothesized.
By introducing the ansatz
\\
\begin{equation}
\frac{\rho_m}{\rho_{\vphi}}=\left(\frac{\rho_m}{\rho_{\vphi}}\right)_{st}\,
+\epsilon  \qquad
\left (\mid \epsilon \mid \ll
\left(\frac{\rho_m}{\rho_{\vphi}}\right)_{st}\right) \, ,
\label{ansatz}
\end{equation}
\\
where the subscript $st$ denotes `stationary', in
\\
\[
\left(\frac{\rho_{m}}{\rho_{\vphi}}\right)^{\displaystyle \cdot} 
= \frac{\rho_m}{\rho_\vphi}
\left[\frac{\dot{\rho}{_m}}{\rho_m} - \frac{\dot{\rho_{\vphi}}}{\rho_{\vphi}}
\right] \, ,
\]
\\
and retaining terms up to first order in the perturbation we get
\\
\begin{equation}
\dot{\epsilon}=3H\left[c^2(r+1)-\frac{1}{r+1}\right]\epsilon \, .
\label{depsilon}
\end{equation}
\\
Hence, the stationary solution will be stable for $c<1/(r+1)$. Given
the currently favored observational data $\rho_m\approx 0.3$ and
$\rho_{\vphi}\approx 0.7$ \cite{expansion} we get $c < 0.7$.

Next we seek an exact solution to Eq. (\ref{consv5}) with the
interaction term given by Eq. (\ref{interaction}) for different
potentials.

$(i)$ For  $V(\vphi(t)) = \beta t^{-m}$,
where $\beta$ and $m$ are positive--definite constants (bearing in mind that
$\rho_m=r\rho_{\vphi}$  and $\rho_{\vphi}=\gamma \,t^{-2}$,
with $\gamma$ a constant), the solution is
\\
\begin{equation}
\vphi(t)=t\; \; _{2}F_{1} \left(\left[\frac{-1}{2},\frac{-1}{2(-2+m)}\right],
\left[\frac{-5+2m}{2(-2+m)}\right];
(\beta/\gamma )^2 \, t^{4-2m} \right) \, ,
\label{vphit1}
\end{equation}
\\
where $_{2}F_{1}$ is the hypergeometric function  \cite{Roach}.
When $m = 2$ the function $_{2}F_{1}$ collapses to
a constant, thereby
\\
\[
\varphi(t)\propto\,t\,\Longrightarrow \,V(\vphi)\propto\,\vphi^{-2}.
\]
\\
This recovers a particular case considered in Ref.\cite{bagla}.

$(ii)$ For $V(\vphi(t)) = \frac{\alpha}{t^{2}}
\left( 1- \frac{\beta}{t^{2}} \right)^{1/2}$ with $\alpha$ and $\beta$ 
positive--definite constants, one finds that
\\
\begin{equation}
\vphi(t) = \beta^{1/2} \ln t \, ,
\label{vphit2}
\end{equation}
\\
where the integration constant has been set to zero. This
potential may look a bit contrived, however, it becomes nearly
exponential, $V(\vphi) \sim e^{-2\vphi}$ (which is another case
considered in \cite{bagla}), for $\beta < 1$.

\section{Concluding remarks}
In this paper we have considered that the present accelerated
expansion of our flat FLRW Universe is driven by an interacting
mixture of cold dark matter and a tachyonic field.  The
interaction was not fixed from the outset but derived from the
requirement that the ratio between the energy densities of both
components remains constant such that there is no coincidence
problem. We have found an exact solution when $\dot{\vphi}^{2} =
{\rm constant}$ and two exact solutions when $\ddot{\vphi} \neq 0$
for specific potentials -Eqs. (\ref{vphit1}) and (\ref{vphit2}).
All these solutions imply power law expansions. For solutions of
this type to exist when $\ddot{\vphi} \neq 0$, a negative scalar
pressure in the matter component is required, in order to keep the
overall equation of state parameter of the cosmic medium constant.

One should be aware that our model is not complete in the sense
that $(i)$ it is unable to provide a dynamical approach towards a
stationary energy density ratio and $(ii)$ it does not include a
radiation dominated era at early times. To achieve this one should 
resort to a more general approach, maybe similar to the one by 
Chimento {\it et al.} \cite{ladw} that extended the scenario of
Ref.\cite{plb} (a scalar field interacting with cold dark matter)
in such a way that both a stationary ratio is dynamically
approached and the radiation era is retrieved for early times.
This, as well as to find solutions other than power law 
expansions will be the subject of future research.

\acknowledgments
This work has been partially supported by the
``Ministerio de Educaci\'{o}n de Chile" through MECESUP Project FSM 9901, the US
A
0108 grant, the Spanish Ministry of Science and Technology under grant
BFM--2003--06033, the ``Direcci\'{o} General de Recerca de Catalunya"
under grant  2001 SGR--00186, and the Deutsche Forschungsgemeinschaft.

\end{document}